\documentclass[english,a4paper,pre,twocolumn,amsmath,amssymb,superscriptaddress,longbibliography]{revtex4-1}
\usepackage{graphicx,color,soul}

\newcommand{\dd}{\mathrm{d}}

\newcommand{\pd}[2]{\frac{\partial #1}{\partial #2}}
\newcommand{\mean}[1]{\langle #1 \rangle}
\newcommand{\Int}[1]{\int\dd #1\;}
\newcommand{\IInt}[3]{\int_{#2}^{#3}\dd #1\;}

\renewcommand{\vec}[1]{\mathbf #1}

\newcommand{\al}{\alpha}
\newcommand{\gam}{\gamma}
\newcommand{\eps}{\varepsilon}

\newcommand{\lam}{\lambda}
\newcommand{\vhi}{\varphi}
\newcommand{\sig}{\sigma}

\newcommand{\id}{\mathbf 1}

\newcommand{\x}{\vec r}

\newcommand{\Dr}{D_\text{r}}
\newcommand{\nois}{\boldsymbol\xi}
\newcommand{\kT}{k_\text{B}T}
\newcommand{\rc}{r_\text{c}}

\begin{document}

\title{Quorum-sensing active particles with discontinuous motility}

\author{Andreas Fischer}
\affiliation{Institut f\"ur Physik, Johannes Gutenberg-Universit\"at Mainz, Staudingerweg 7-9, 55128 Mainz, Germany}
\affiliation{Graduate School Materials Science in Mainz, Staudinger Weg 9, 55128 Mainz, Germany}
\author{Friederike Schmid}
\affiliation{Institut f\"ur Physik, Johannes Gutenberg-Universit\"at Mainz, Staudingerweg 7-9, 55128 Mainz, Germany}
\author{Thomas Speck}
\affiliation{Institut f\"ur Physik, Johannes Gutenberg-Universit\"at Mainz, Staudingerweg 7-9, 55128 Mainz, Germany}

\begin{abstract}
  We develop a dynamic mean-field theory for polar active particles that interact through a self-generated field, in particular one generated through emitting a chemical signal. While being a form of chemotactic response, it is different from conventional chemotaxis in that particles discontinuously change their motility when the local concentration surpasses a threshold. The resulting coupled equations for density and polarization are linear and can be solved analytically for simple geometries, yielding inhomogeneous density profiles. Specifically, here we consider a planar and circular interface. Our theory thus explains the observed coexistence of dense aggregates with an active gas. There are, however, differences to the more conventional picture of liquid-gas coexistence based on a free energy, most notably the absence of a critical point. We corroborate our analytical predictions by numerical simulations of active particles under confinement and interacting through volume exclusion. Excellent quantitative agreement is reached through an effective translational diffusion coefficient. We finally show that an additional response to the chemical gradient direction is sufficient to induce vortex clusters. Our results pave the way to engineer motility responses in order to achieve aggregation and collective behavior even at unfavorable conditions.
\end{abstract}

\maketitle


\section{Introduction}

Motility is a fundamental property of many biological systems, from swimming sperm~\cite{alvarez14} and crawling cells~\cite{abercrombie80} to animal locomotion~\cite{sponberg17}. While animals coordinate motion in response to optical and mechanical sensory inputs, regulation of motility on the microscale requires very different mechanisms. A particular challenge is to overcome the physical limitations of sensing and gathering information about the environment on small scales~\cite{bialek12}.

For example, sperm cells can detect and follow gradients of signaling molecules (a small peptide in the case of sea urchin) through changing the beat pattern of their flagellum~\cite{kaupp03,friedrich07}. An alternative strategy to gather information about the local environment is quorum sensing, in which members of a population exude signaling molecules~\cite{miller01}. The sensed concentration (or rather concentration change) influences gene expression and determines the phenotype. Quorum sensing facilitates the synchronized behavior of bacteria required, \emph{e.g.}, for biofilm formation~\cite{parsek07}, virulence~\cite{zhu02}, bioluminescence~\cite{lupp05}, and motility control~\cite{sperandio02}. Moreover, there is evidence that quorum sensing also plays a role in the regulation of immune cell responses~\cite{antonioli19}. Beyond the understanding of large-scale collective behavior of biological entities, such communication mechanisms are interesting for synthetic microrobots with a limited set of responses. Even if individual robots can only follow simple rules, there have been studies recently reporting on the emerging collective behavior in ensembles of robots with the potential to fulfill complex tasks~\cite{rubenstein14,scholz18,yu18,xie19,li19}. Another possible application is to guide the spontaneous assembly of active functional materials~\cite{aubret18}.

Phoretic colloidal particles provide a well-characterized experimental model system in which we can study different aspects of emerging collective behavior~\cite{bech16}. Implementing responses beyond excluded volume and alignment requires control over particles motion~\cite{dai16,mano17,khadka18}. For phoretic Janus particles triggered by light, individual control of motility has been demonstrated recently for two modes of perception: cohesive flocking through vision cones~\cite{lavergne19} and for quorum sensing~\cite{bauerle18}. For the later, a virtual concentration field is calculated from particle positions that determines the propulsion speed of each particle through a feedback mechanism. Switching between motile and passive according to a preset threshold, aggregation into a cluster is observed, the shape and density of which can be controlled through tailoring the response of particles.

Such aggregation of spherical active particles into dense clusters is often rationalized within the framework of motility-induced phase separation~\cite{cate15}, which is based on an effective propulsion speed $v(\rho)$ that is reduced as the local density $\rho$ increases. The system becomes inhomogeneous with coexisting dilute and dense domains~\cite{yaou12,redn13,butt13,wysocki14}. Theory shows that this phase separation can be understood as a large-scale dynamic instability captured through non-linear evolution equations~\cite{bial13,spec15}. It requires the derivative of $v(\rho)$ and thus a continuous function. While in principle a continuous response might be achieved through quorum sensing, in realistic applications as mentioned above the response is \emph{discontinuous}.

Here we develop a comprehensive theoretical framework for this scenario, in which active particles interact through a self-generated chemical field. In particular, we consider discontinuous changes of particle motility (another possibility would be the diffusion coefficient~\cite{liu11,fu12}). Somewhat counter-intuitive, such discontinuous changes simplify the mathematical problem since the resulting evolution equations are now linear, which allows us to obtain analytical results for two simple geometries: a planar interface and circular clusters. Our theory is a mean-field theory neglecting fluctuations. However, we argue that the dominating effect of fluctuations is to increase the diffusive current out of clusters. Taking this effect into account, we achieve excellent quantitative agreement of our theoretical predictions with simulations of interacting active particles. In the final part we consider vortex clusters, which in active matter are typically induced by a combination of aligning inter-particle interactions and confinement~\cite{gross08,wiol13,bric15}, or are related to ``active turbulence''~\cite{thampi14,james18}. Here we show that similar patterns can be achieved through an additional response of particle orientations to the chemical field gradient.


\section{Model}

\subsection{Basic equations}

\begin{figure}[t]
  \centering
  \includegraphics{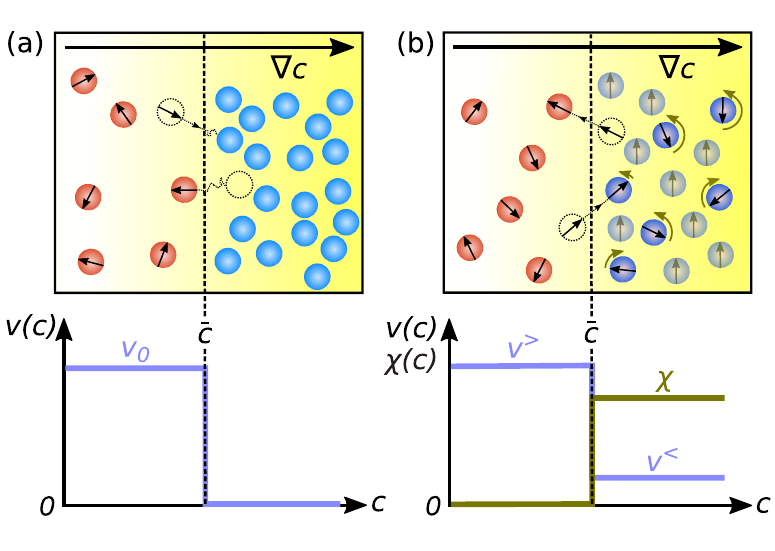}
  \caption{Discontinuous particle response to a self-produced chemical concentration field $c(\x,t)$ (yellow). (a) Motility response $v(c)$: Particles propel with velocity $v_0$ when they sense concentrations smaller than the threshold $\bar{c}$ and are passive ($v=0$) when their measured concentration surpasses $\bar{c}$. This leads to the formation of an active and a passive domain with the boundary set by $c(\x,t)=\bar{c}$. (b) Motility and orientational response $\chi(c)$: Particles switch from high speed $v^>$ to small speed $v^<$ and orient perpendicular to $\nabla c$ when exceeding $\bar{c}$.}
  \label{fig:sketch}
\end{figure}

We consider a suspension of polar active particles in two dimensions governed by the stochastic equations of motion
\begin{equation}
  \label{eq:eom}
  \dot\x = v\vec e + \nois, \qquad \dot\vhi = \vec e\cdot\chi\vec s + \eta.
\end{equation}
Each particle is described by a unit orientation $\vec e\equiv(\cos \vhi,\sin \vhi)^T$ (with $\vhi$ the angle enclosed between the orientation and the $x$-axis) along which it is propelled with speed $v$. Here, $\nois$ and $\eta$ are translational and orientational Gaussian noise with zero mean and diffusion coefficients $D_0$ and $\Dr$, respectively. With these parameters, we define the typical length and speed scale,
\begin{equation}
  \ell \equiv \sqrt{D_0/\Dr}, \qquad v_\ast \equiv 4\sqrt{D_0\Dr} = 4D_0/\ell.
\end{equation}
The orientation of particles aligns perpendicular to the direction $\vec s$ with strength $\chi$. Both the speed $v(c)$ and the coupling $\chi(c)$ are functions of the local concentration $c(\x,t)$ of some chemical that is produced or consumed by the particles, see Fig.~\ref{fig:sketch}. This model focuses on the interactions mediated by the chemical field and neglects direct interactions (\emph{e.g.} through volume exclusion) but also phoretic and hydrodynamic interactions.

Previous studies of related models~\cite{pohl14,saha14,lieb15,stark18,lieb18} have focused on chemotactic behavior in which alignment occurs parallel to $\nabla c$, describing either chemo-attractive or chemo-repellent particles depending on the sign of $\chi$. Setting $v\vec e\propto\nabla c$ has been used to model the dynamics of chemically active uniform particles~\cite{soto14,niu18,saha19}. In contrast, here we will consider the case of perpendicular alignment to the concentration gradient ($\vec s\propto\nabla c$). More importantly, we will consider discontinuous functions $v(c)$ and $\chi(c)$ that switch between two constant values depending on a threshold $\bar c$ (as depicted in Fig.~\ref{fig:sketch}).

An equivalent but for our purposes more convenient representation of Eq.~\eqref{eq:eom} is through the evolution of the joint probability $\psi(\x,\vhi;t)$ of position and orientation, which reads
\begin{equation}
  \label{eq:psi}
  \pd{\psi}{t} = -\nabla\cdot[v\vec e-D_0\nabla]\psi - \chi\pd{}{\vhi}[\vec e\cdot\vec s\psi] + \Dr\pd{^2\psi}{\vhi^2}.
\end{equation}
Decomposing $\psi(\x,\vhi;t)$ into Fourier modes leads to a hierarchy of coupled dynamic equations~\cite{bertin06}. Since we are interested in the large-scale collective behavior, we close the hierarchy at the second level and treat the system in terms of the first two moments of the joint probability, the particle density $\rho(\x,t)$ and the polarization $\vec p (\x,t)$. The derivation of their evolution equations can be found in appendix~\ref{sec:hydro}. The equation for the density is the continuity equation
\begin{equation}
  \label{eq:rho}
  \pd{\rho}{t} = -\nabla\cdot\vec j
\end{equation}
with particle current
\begin{equation}
  \label{eq:j}
  \vec j = v\vec p - D_0\nabla\rho.
\end{equation}
The total current is the sum of an active current due to a non-zero polarization and a diffusive current due to density gradients.

The general expression for the polarization is a bit lengthy and reads
\begin{multline}
  \label{eq:p}
  \pd{\vec p}{t} = \nabla\cdot\vec M + \frac{1}{2}\chi\rho\vec R\cdot\vec s - \left(\Dr+\frac{\chi^2}{8\Dr}|\vec s|^2\right)\vec p \\
  + \frac{v}{16\Dr}\left[(\nabla\cdot\chi\vec s)\vec R\cdot\vec p-(\nabla\cdot\vec R\cdot\chi\vec s)\vec p\right],
\end{multline}
where the divergence involves the matrix
\begin{multline}
  \label{eq:M}
  \vec M \equiv -\frac{1}{2}v\rho\id + D_0\nabla\vec p + \frac{D_0v}{v_\ast^2}\left[(\nabla v\vec p)+\vec R\cdot(\nabla v\vec p)\cdot\vec R\right] \\
  + \frac{\chi v}{16\Dr}\left[3(\vec s\vec p)\cdot\vec R-\vec R\cdot(\vec s\vec p)\right]
\end{multline}
akin to the stress in conventional hydrodynamics. It is convenient to employ the asymmetric matrix
\begin{equation}
  \vec R \equiv \left(\begin{array}{cc} 0 & -1 \\ 1 & 0\end{array}\right)
\end{equation}
describing a rotation by $\pi/2$.

We assume that the chemicals are produced uniformly by every particle with rate $\gam$ and diffuse independently with diffusion coefficient $D_\text{c}$ in the semi-space above the plane in which the active particles are moving. Moreover, assuming a separation between the time scale on which active particles move and the time scale for the chemical profile to relax, one has
\begin{equation}
  \label{eq:c}
  c(\x) = \frac{\gam}{4\pi D_\text{c}}\Int{^2\x'} \rho(\x-\x')\frac{e^{-r'/\lam}}{r'}
\end{equation}
with decay length $\lam$ due to the degradation of chemicals and $r'=|\x'|$. For an unbounded system with uniform density $\rho_0$, we find the uniform concentration
\begin{equation}
  \label{eq:c_uni}
  c_0 = \frac{\gam\lam}{2D_\text{c}}\rho_0.
\end{equation}

\subsection{Active Brownian particles}

For $\chi=0$ and a linear relationship $v(c)=v_0-\zeta c$ between speed and local concentration, our model reduces to that of conventional active Brownian particles. For densities that vary only weakly on the interaction range $\lam$, we can expand $\rho(\x-\x')\approx\rho(\x)-\x'\cdot\nabla\rho+\frac{1}{2}\x'^2\nabla^2\rho$, leading to
\begin{equation}
  v(\x,t) = v_0 - \frac{\zeta\gam\lam}{2D_\text{c}}\rho(\x,t) - \frac{\zeta\gam\lam^3}{2D_\text{c}}\nabla^2\rho(\x,t)
\end{equation}
so that the effective speed indeed depends on the local density. In contrast, here we go beyond such a simple analytical relation and study situations where the speed $v$ and the coupling $\chi$ change \emph{discontinuously} depending on the chemical field $c$ reaching a threshold $\bar c$ [Fig.~\ref{fig:sketch}(a)].

\subsection{Continuity condition}

\begin{figure}[t]
  \centering
  \includegraphics{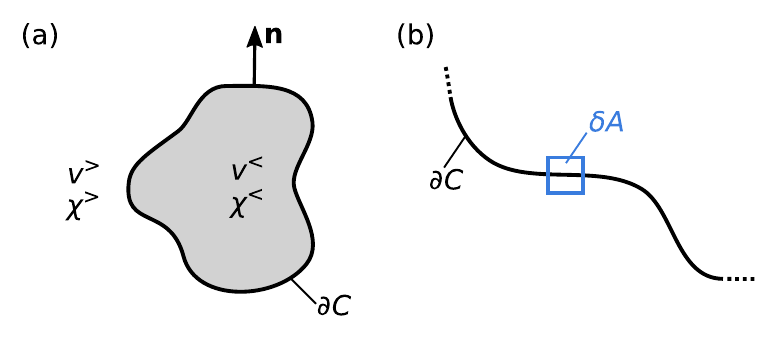}
  \caption{Continuity condition. (a)~Domain $\mathcal C$ inside which $c>\bar c$ bounded by the curve $\partial\mathcal C$ defined through $c=\bar c$. Along this curve, the speed $v(c)$ and the coupling $\chi(c)$ are discontinuous. (b)~Zoom to part of the boundary curve $\partial C$ with integration area $\delta A$.}
  \label{fig:cont}
\end{figure}

Drawing an analogy with electrostatics, the speed (and coupling) is akin to a dielectric constant that jumps at the interface between different media. Consequently, the discontinuity of $v$ and $\chi$ imposes jump conditions for $\vec p (\x)$ at $\partial\mathcal C$ via Eq.~\eqref{eq:p}, which we derive in analogy to the jump conditions for the electromagnetic fields. To this end, we assume that a closed domain $\mathcal C$ has formed with speed $v^<$ and coupling $\chi^<$ inside, and $v^>$ and $\chi^>$ outside [Fig.~\ref{fig:cont}(a)]. The domain is bounded by the curve $\partial\mathcal C$ with normal vector $\vec n$ pointing outwards. In the steady state ($\partial_t\vec p=0$), integrating Eq.~\eqref{eq:p} over an infinitesimal area $\delta A$ with one half inside and the other half outside $\mathcal C$ [cf. Fig.~\ref{fig:cont}(b)] yields
\begin{equation}
  \IInt{^2\x}{\delta A}{} \nabla\cdot\vec M + \IInt{^2\x}{\delta A}{}\frac{v}{16\Dr}(\nabla\cdot\chi\vec s)\vec R\cdot\vec p = 0.
\end{equation}
The first integral can be evaluated using the divergence theorem. The second integral involves the gradient of $\chi$, which jumps at $\x=\x_\text{c}$ (with $\x_\text{c}$ a parametrization of $\partial\mathcal C$) and is given by
\begin{equation}
  \nabla\chi = (\chi^>-\chi^<)\vec n\delta(\x-\x_\text{c}).
\end{equation}
The speed $v$ also jumps at $\x_c$ and we follow Ref.~\citenum{griffiths99} to evaluate $\delta(\x-\x_\text{c})$, by which we arrive at the jump condition
\begin{multline}
  \label{eq:cont_cond}
  \frac{1}{16\Dr}(\vec s\cdot\vec n)(\chi^>-\chi^<)\frac{1}{2}(v^>\vec R\cdot\vec p^>-v^<\vec R \cdot \vec p^<) \\ +\vec n \cdot (\vec M^>-\vec M^<) = 0
\end{multline}
with ($\vec{M}^<$ and $\vec{M}^>$ are defined analogously)
\begin{equation}
  \vec p^< = \lim\limits_{|\boldsymbol \eps| \rightarrow 0}{\vec{p}(\x_\text{c}-\boldsymbol \eps)}, \qquad
  \vec p^> =\lim\limits_{|\boldsymbol  \eps| \rightarrow 0}{\vec{p}(\x_\text{c}+\boldsymbol \eps)}.
\end{equation}
The continuity equation~\eqref{eq:rho} determines whether the density is continuous ($D_0>0$) or also becomes discontinuous ($D_0=0$, see next Sec.~\ref{sec:pol}). Note that in any case, the concentration profile Eq.~\eqref{eq:c} remains a continuous function everywhere.

\section{No alignment}

\subsection{Polarization}
\label{sec:pol}

We now study the conditions under which particles aggregate and form a stationary inhomogeneous system. We first simplify the general problem and assume that there is no alignment of orientations with the field gradient through setting $\chi=0$, cf. Fig.~\ref{fig:sketch}(a). The speed $v$ jumps across the interface but is constant within each region with
\begin{equation}
  \label{eq:noa:M}
  \vec M_0 = -\frac{1}{2}v\rho\id+D_0\left(1+\frac{v^2}{v_\ast^2}\right)\nabla\vec p
\end{equation}
and the evolution of the polarization is thus governed by
\begin{equation}
  \pd{\vec p}{t} = -\frac{v}{2}\nabla\rho + D_0\left(1+\frac{v^2}{v_\ast^2}\right)\nabla^2\vec p - \Dr\vec p.
\end{equation}
In the following, we exploit that in the steady state (with no-flux boundary conditions) not only the divergence but the particle current vanishes, $\vec j=0$. We then substitute $\nabla\rho=(v/D_0)\vec p$ [Eq.~\eqref{eq:j}] to obtain a differential equation for the polarization $\vec p$ alone. Clearly, $\vec p=0$ is a solution, which corresponds to a homogeneous system.

Now let us assume that a cluster has formed due to a reduced speed $v^<$. The continuity condition Eq.~\eqref{eq:cont_cond} simplifies to
\begin{equation}
  \label{eq:noa:cont}
  \vec n \cdot (\vec M_0^>-\vec M_0^<) = 0.
\end{equation}
If we set $D_0=0$ then from Eq.~\eqref{eq:j} we immediately find $\vec p=0$ everywhere and thus $v\rho=\text{const}$. This relationship has been exploited to create complex density patterns through spatially modulating the speed~\cite{sten16,fran18}. If $v^<=0$ then $\rho^>=0$, \emph{i.e.}, all particles condensate into the cluster and stop moving. In the following, we consider the coexistence of motile and passive particles with non-vanishing diffusion coefficient $D_0>0$.

\subsection{Planar interface}
\label{sec:planar}

We first assume a geometry in which the interface is planar with normal vector $\vec n=\vec e_x$. The system is translationally invariant along the transversal direction so that $\rho(x)$ only depends on $x$ and the polarization is $\vec p=p(x)\vec e_x$. For $x<0$ we set $v^<=0$ implying a constant density $\rho_\text{c}$ and $p=0$. For $x>0$ with $v^>=v_0$ we find the differential equation $p''-p/\xi_0^2=0$ (the prime denotes the derivative) with decay length
\begin{equation}
  \label{eq:xi}
  \xi_0 \equiv \ell\left[\frac{1+(v_0/v_\ast)^2}{1+8(v_0/v_\ast)^2}\right]^{1/2},
\end{equation}
which equals $\ell$ for $v_0=0$ and drops to $\ell/\sqrt8$ for $v_0\to\infty$. The solution is $p(x)=ae^{-x/\xi_0}$ with integration constant $a$. The polarization thus has a discontinuity at $x=0$, where it jumps from $a<0$ to zero. The density follows through integrating Eq.~\eqref{eq:j} as
\begin{equation}
\label{eq:rho_planar}
  \rho(x) = \rho_\text{c} - \frac{v_0\xi_0}{D_0} a \left(e^{-x/\xi_0}-1\right)
\end{equation}
for $x\geqslant 0$ and $\rho=\rho_\text{c}$ for $x<0$. For $x\gg \xi_0$ the active gas reaches $\rho_\text{g}=\rho_\text{c}+\frac{v_0\xi_0}{D_0} a$. The continuity condition~\eqref{eq:noa:cont} yields
\begin{equation}
  \label{eq:cont_planar}
  0 = \frac{1}{2}v_0\rho_\text{c} + D_0\left(1+\frac{v_0^2}{v_\ast^2}\right)\frac{a}{\xi_0},
\end{equation}
which relates $a$ to the density $\rho_\text{c}$. Equation~\eqref{eq:cont_planar} can be rewritten into
\begin{equation}
  \label{eq:cont_planar2}
  \frac{\rho_\text{g}}{\rho_\text{c}} = \left(1+8\frac{v_0^2}{v_\ast^2}\right)^{-1}
\end{equation}
showing that the ratio of gas and cluster density drops quadratically with the propulsion speed.

\begin{figure}[t]
  \centering
  \includegraphics{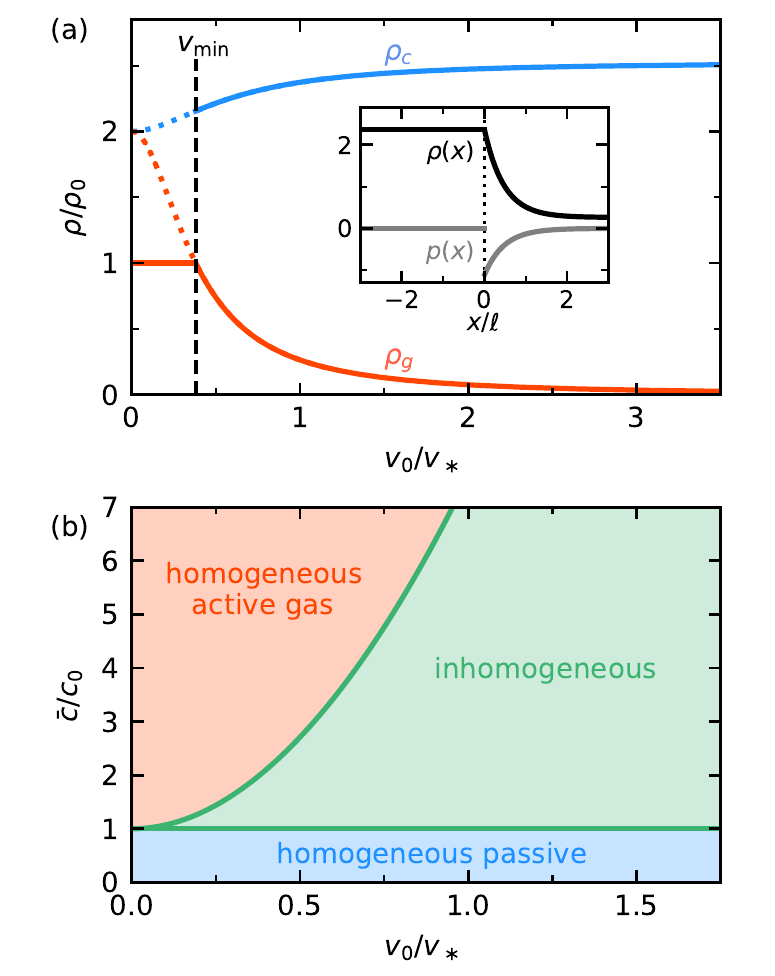}
  \caption{Planar interface. (a)~Passive ($\rho_\text{c}$) and active gas ($\rho_\text{g}$) densities as a function of propulsion speed for $\bar c/c_0=2$ and $\lam/\ell=0.45$ with respect to a uniform system at density $\rho_0$. There is a minimal speed $v_\text{min}$ below which the system remains a homogeneous active gas. The inset shows the profiles $\rho(x)$ and $p(x)$ at $v_0/v_\ast=1$. (b)~Corresponding $v_0$--$\bar{c}$ phase diagram. As the threshold $\bar{c}$ is increased, $v_\text{min}$ rises (upper green line). For thresholds below the concentration of the uniform system $c_0$, another homogeneous region exists in which all particles become passive.}
  \label{fig:planar}
\end{figure}

The final condition reads $c(x=0)=\bar c$ with concentration
\begin{multline}
  c(x=0) = \\ \frac{\gam}{4\pi D_\text{c}} \IInt{r}{0}{\infty} e^{-r/\lam} \left[\pi\rho_\text{c} + \IInt{\theta}{0}{\pi}\rho(r\sin\theta)\right]
\end{multline}
at the interface. The integrals can be performed analytically, which yields the closed expression
\begin{equation}
  \label{eq:cBar_planar}
  \bar c = \frac{\gam\lam}{2D_\text{c}}\left[\rho_\text{c}+\frac{v_0\xi_0}{2D_0}a\left(1-\frac{2}{\pi}\frac{\cos^{-1}(\lam/\xi_0)}{\sqrt{1-(\lam/\xi_0)^2}}\right)\right].
\end{equation}
For a given threshold $\bar{c}$ and decay length $\lam$, Eqs.~\eqref{eq:rho_planar}, \eqref{eq:cont_planar}, and \eqref{eq:cBar_planar} can be solved for the passive and active gas densities $\rho_\text{c}$ and $\rho_\text{g}$, which are plotted Fig.~\ref{fig:planar}(a) as a function of the reduced outer speed $v_0/v_\ast$. The coexistence of a passive and an active domain is possible for speeds $v_0$ higher than a minimal speed $v_\text{min}$ which is set by the condition $\rho_\text{g}=\rho_0$. For smaller $v_0$, the system remains a homogeneous active gas with $\rho(x)=\rho_0$. With rising speed, the cluster density $\rho_\text{c}$ saturates at
\begin{equation}
  \label{eq:rho_c}
  \rho_\text{c}(v_0\to\infty) = \bar\rho\left[\frac{1}{2}+\frac{\cos^{-1}(\sqrt 8\lam/\ell)}{\pi\sqrt{1-8(\lam/\ell)^2}}\right]^{-1}
\end{equation}
with $\bar\rho\equiv\frac{2D_\text{c}}{\gam\lam}\bar c$, while the gas density $\rho_\text{g}$ converges to zero due to the enhanced active current into the passive region.

\subsection{Coexistence without a critical point}

We now gather our results to produce the phase diagram shown in Fig.~\ref{fig:planar}(b) varying speed $v_0$ and threshold $\bar c$. There are two homogeneous phases in which all particles are either motile (forming an active gas with concentration $c_0<\bar c$) or passive ($c_0>\bar c$). Between these homogeneous phases, the system becomes spatially inhomogeneous and develops an interface between a dense passive and a dilute active region. At greater thresholds $\bar{c}$, a higher speed $v_\text{min}$ is required to enable phase coexistence.

\begin{figure}[b!]
  \centering
  \includegraphics{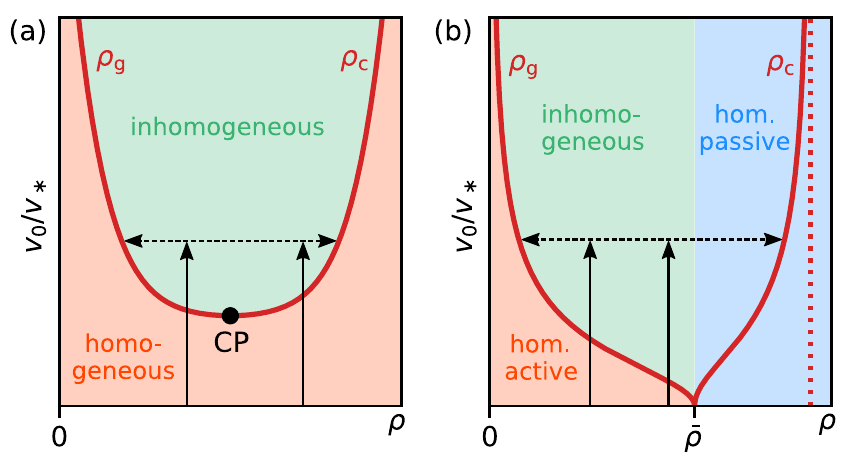}
  \caption{Paradigms for inhomogeneous systems. (a)~Conventional phase diagram for liquid-gas phase separation as observed for active Brownian particles. The two-phase region is bounded by the binodal (thick line), which terminates in a critical point (CP). Arrows show two global densities yielding the same coexisting densities (but different relative size of dense to dilute regions). (b)~Phase diagram for discontinuous motility response at fixed threshold $\bar c$. Notable is the absence of a critical point. For $\rho_0>\bar\rho$, the system becomes homogeneous and passive. The dashed line is the prediction Eq.~\eqref{eq:rho_c} for the maximal cluster density.}
  \label{fig:coex}
\end{figure}

Such a coexistence of different densities is a hallmark of passive liquid-gas phase separation (at constant volume), which requires attractive interactions between particles. Coexistence is then related to the equality of the chemical potential in both phases. It has been shown that the mean-field behavior of active Brownian particles can be mapped onto an effective free energy~\cite{spec14,spec15} and thus follows the same scenario as passive liquid-gas coexistence, although the non-equilibrium nature manifests itself in phenomena like a negative interfacial tension~\cite{bial15} and ``bubbly'' phase separation~\cite{tjhung18}. Nevertheless, the generic relation with passive liquid-gas coexistence has been corroborated by detailed numerical investigations of the phase behavior~\cite{redn13,bial15,sieb17,levis17,sieb18}. The resulting phase diagram in the $v_0$--$\rho$ plane is sketched in Fig.~\ref{fig:coex}(a). Quenching the system inside the two-phase region, the stable steady state is inhomogeneous with the coexisting densities given by the points on the binodal. Changing the global density changes the relative size of the regions covered by either phase as expressed by the lever rule, but not the coexisting densities. Phase coexistence is terminated in a critical point characterized by diverging fluctuations below which the system remains homogeneous at any density.

For the discontinuous motility studied here, we also reach a state of coexistence. As for conventional liquid-gas coexistence, varying the global density $\rho_0$ yields the same coexisting densities. There are, however, important differences [Fig.~\ref{fig:coex}(b) for fixed threshold $\bar c$]. First, for global densities $\rho_0>\bar\rho$ the full system becomes homogeneous and passive since $c_0>\bar c$. For coexistence, we need to cross the gas density, $v_0>v_\text{min}$, with $\rho_\text{c}>\bar\rho$. Second, there is no critical point. The two-phase region becomes more narrow as the speed $v_0$ is decreased and ends in a single point at $v_0=0$ with density $\bar\rho$ [cf. Eq.~\eqref{eq:cont_planar2}]. This feature enables stable aggregation even at very low speeds in a region of parameter space that is mainly controlled by the threshold $\bar c$.

\subsection{Circular clusters}
\label{seq:circular}

We now assume a circular cluster of radius $\rc$ with normal vector $\vec n=\vec e_r$ of the bounding curve $\partial\mathcal C$. In steady state, the current $\vec j=0$ still vanishes and we thus anticipate a radially symmetric density profile $\rho(r)$ so that the polarization $\vec p(r)=p(r)\vec e_r$ only has a radial component. Going to polar coordinates (with $r=0$ the center of mass), we obtain the modified Bessel differential equation
\begin{equation}
  \label{eq:pr}
  0 = p'' + \frac{p'}{r} - \frac{p}{r^2} - \frac{p}{\xi_0^2}
\end{equation}
with the length scale $\xi_0$ defined in Eq.~\eqref{eq:xi}. The general solution of Eq.~\eqref{eq:pr} reads
\begin{equation}
  p(r) = a_1I_1(r/\xi_0) + a_2K_1(r/\xi_0),
\end{equation}
where $I_n(x)$ and $K_n(x)$ are modified Bessel functions of order $n$ of the first and second kind, respectively, and $a_1$ and $a_2$ are integration constants to be determined. In the outer region, $p\to 0$ as $r\to\infty$ and thus $a_1^>=0$. The density profile is obtained through integrating $\rho'=(v_0/D_0)p$ [Eq.~\eqref{eq:j}],
\begin{equation}
  \label{eq:noa:rho}
  \rho(r) = \frac{v_0\xi_0}{D_0}\left[a_1I_0(r/\xi_0)-a_2K_0(r/\xi_0)\right] + b
\end{equation}
with another integration constant $b$. For the integral of the density (\emph{viz.} number of particles) to vanish as $\rc\to0$ requires to set $a_2^<=0$. Moreover, $b^>=\rho_\text{g}$ since the density is that of the active gas for $r\to\infty$.

We now consider the formation of clusters through turning off the self-propulsion when triggering the threshold ($v^<=0$), which is the situation realized experimentally in Ref.~\citenum{bauerle18}. As for the planar interface, this implies $p=0$ and a constant density $\rho_\text{c}$ inside the cluster. The density profile is thus fixed by $\rho_\text{c}$, $\rho_\text{g}$, $\rc$, and $a_2^>$, which we need to determine.

To this end, the first condition follows from the continuity of the density at $\rc$,
\begin{equation}
  \rho_\text{c} = -\frac{v_0\xi_0}{D_0}a_2^>K_0(\rc/\xi_0) + \rho_\text{g}.
\end{equation}
To fulfill the continuity condition~\eqref{eq:noa:cont}, we need the expression
\begin{equation}
  \vec e_r\cdot\vec M_0 = \left[-\frac{1}{2}v\rho + D_0\left(1+\frac{v^2}{v_\ast^2}\right)p'\right]\vec e_r,
\end{equation}
where the derivative of the polarization can be expressed as (inside the cluster with $b^<=\rho_\text{c}$)
\begin{equation}
  p' = \frac{D_0}{v\xi_0^2}(\rho-b) - \frac{p}{r}.
\end{equation}
Our second condition then follows through the continuity condition, which can be recast into
\begin{equation}
  \label{eq:noa:cont_rad}
  \rho_\text{c} = \left(1+8\frac{v_0^2}{v_\ast^2}\right)\rho_\text{g} + \frac{v_0}{\Dr}\left(1+\frac{v_0^2}{v_\ast^2}\right)\frac{p_\text{c}}{\rc}.
\end{equation}
The polarization jumps from $p=0$ inside the cluster to 
\begin{equation}
  p_\text{c} = a_2^>K_1(\rc/\xi_0) < 0
\end{equation}
at $\rc$. As expected, for a cluster of infinite size ($\rc\to\infty$), Eq.~\eqref{eq:noa:cont_rad} becomes equal to the result for the planar interface, Eq.~\eqref{eq:cont_planar2}.

The concentration inside the cluster needs to exceed the threshold $\bar c$ with $c(\rc)=\bar c$ at the boundary, which constitutes our third condition. Plugging the density profile Eq.~\eqref{eq:noa:rho} into Eq.~\eqref{eq:c}, the condition becomes
\begin{equation}
  \label{eq:cBar_rad}
  \frac{\bar c}{c_0} = \IInt{r'}{0}{\infty} r'\frac{\rho(r')}{\rho_0} \IInt{k}{0}{\infty} \frac{k}{\sqrt{1+(k\lam)^2}} J_0(k\rc)J_0(kr') 
\end{equation}
with $J_0(x)$ the Bessel function of first kind and order zero. No closed analytical expression is available for these integrals but they can be solved easily numerically.

For the remainder of this paper, we adopt as specific geometry a circular confinement of radius $R$ containing a single cluster with the center of mass at the center of the confinement. This geometry has been studied previously in Ref.~\citenum{rein16} but for run-and-tumble dynamics without translational diffusion. Our fourth and final condition then follows through the conservation of particle number,
\begin{equation}
  \IInt{^2\x}{}{}\rho(\x) = \rho_0 A,
\end{equation}
where $A=\pi R^2$ is the area of the system. We thus have four conditions for four unknowns, which we solve iteratively for given threshold $\bar c$ and decay length $\lam$. The minimal threshold above which a passive cluster can coexist with an active gas is given by $\bar{c}_\text{min}=c(R)$ and can be calculated analytically from Eq.~\eqref{eq:cBar_rad} as
\begin{equation}
  \label{eq:cMin}
  \frac{\bar{c}_\text{min}}{c_0}=\frac{1}{2}\left[1-I_0(2R/\lam)+L_0(2R/\lam)\right]
\end{equation}
with $L_0(x)$ the modified Struve-function of order zero.

\begin{figure*}[t]
  \centering
  \includegraphics{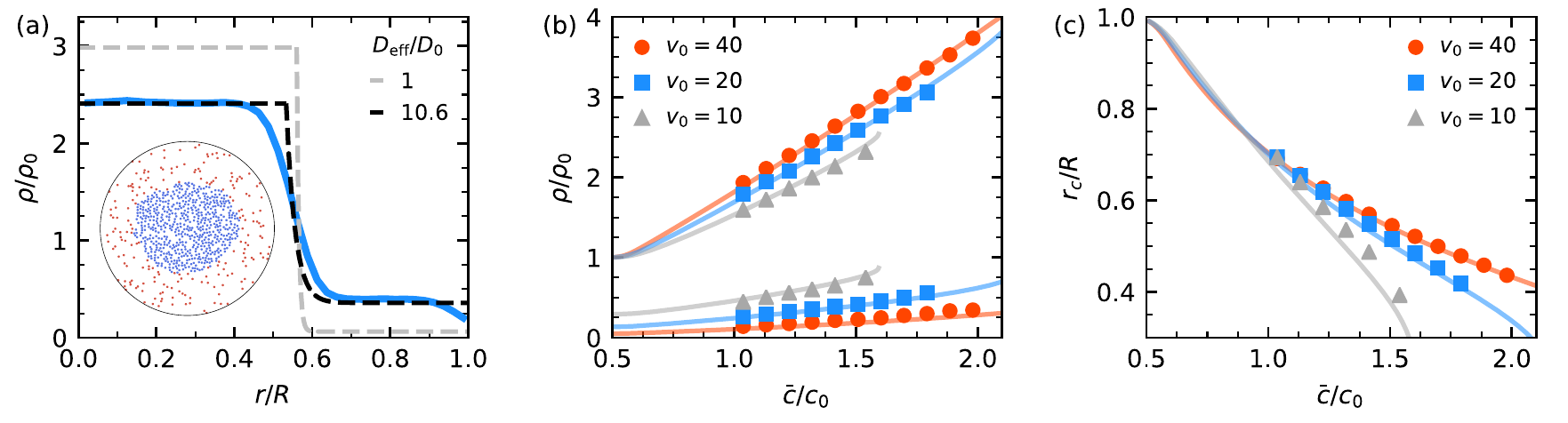}
  \caption{Circular clusters. (a)~Comparison of radial density profile $\rho(r)$ from simulations (blue) and theory [Eq.~\eqref{eq:noa:rho}]: unmodified (gray line) and with effective diffusion coefficient $D_\text{eff}=10.6D_0$ (black line) for $v_0=20$, $\lam=10$, $\bar c=1.4c_0$. The inset shows a corresponding simulation snapshot with a passive cluster (blue) surrounded by an active gas (red). (b)~Cluster and gas densities versus threshold $\bar{c}$ at $\lam=10$ for different propulsion speeds $v_0$ (symbols: simulations, lines: theory with fitted $D_\text{eff}$). At higher $\bar{c}$ we do not observe the formation of stable clusters. Clusters also exist at lower $\bar{c}$ than shown here, but in that case the gas density is not measurable reliably due to the influence of the boundary. (c)~Cluster size $\rc$ versus $\bar{c}$ for different $v_0$ [same parameters as in (b)]. Standard errors are smaller than symbol size.}
  \label{fig:circular}
\end{figure*}

\subsection{Simulations} 
\label{seq:noa:simulations}

\subsubsection{Simulation details}

To test our analytical predictions we now turn to numerical simulations. We numerically solve Eq.~\eqref{eq:eom} with $\chi=0$ for $N=800$ particles additionally interacting via the repulsive Weeks-Chandler-Anderson (WCA) potential~\cite{week72}
\begin{equation}
  \label{eq:wca}
  u(r_{ij}) =
  \begin{cases}
    4\eps\left[\left(\frac{\sig}{r_{ij}}\right)^{12} - \left(\frac{\sig}{r_{ij}}\right)^6 
      + \frac{1}{4}\right] & (r_{ij}/\sig < 2^{1/6}) \\
    0 & (r_{ij}/\sig \geq 2^{1/6}),
  \end{cases}
\end{equation}
with particle separation $r_{ij}=|\x_j-\x_i|$ and $\x_i$ the position of the $i$th particle. We use a potential strength $\eps=100\kT$ modeling hard discs with an effective diameter $\sig_\text{eff}=1.10688\sig$~\cite{bark67}. The rotational diffusion coefficient is set to $\Dr=3D_0/\sig_\text{eff}^2$. The same set of parameters has been used in previous studies of active Brownian particles~\cite{sieb17,sieb18}. From now on, we employ $\sig$, $\sig^2/D_0$, and $k_B T$ as units of length, time, and energy, respectively. The particles move in a circular confinement with radius $R=42.4$ at packing fraction $\phi=N \frac{(\sig_\text{eff}/2)^2}{R^2}=0.136$ (corresponding to particle number density $\rho_0=0.141$). To prevent the accumulation of particles at the boundary (due to their persistent motion~\cite{yaou14,smallenburg15,caprini18b}), particle orientations are flipped instantaneously and redirected towards the center of the confinement when reaching the border.

Every particle acts as a point source of the chemical, thus the instantaneous concentration sensed by particle $i$ is
\begin{equation}
  c_i \equiv c(\x_i) = \frac{\gam}{4\pi D_\text{c}}\sum_{j \neq i}^N \frac{e^{-r_{ij}/\lam}}{r_{ij}}.
\end{equation}
When $c_i$ is larger than the threshold value $\bar c$, the particle's propulsion speed is set to zero. 

For different decay lengths $\lam$, we perform simulations varying thresholds $\bar c$ and propulsion speeds $v_0$ at which the system separates into a dense cluster of passive particles surrounded by a dilute gas of active particles, see snapshot in Fig.~\ref{fig:circular}(a). After the system has reached the steady state, we measure the radial density profile $\rho(r)$ with respect to the center of mass of all particles. Density profiles can be well fitted by
\begin{equation}
  \rho(r) = \frac{\rho_\text{c}+\rho_\text{g}}{2}+\frac{\rho_\text{c}-\rho_\text{g}}{2}\tanh\left(\frac{r-\rc}{2\omega}\right),
\end{equation}
from which we obtain the coexisting densities $\rho_\text{c}$ and $\rho_\text{g}$, the cluster size $\rc$, as well as the width of the interface $\omega$. Both $\rho_c$ and $\rho_g$ increase with rising threshold $\bar{c}$, while the cluster radius decreases, cf. Fig.~\ref{fig:circular}(b,c), \emph{i.e.}, the cluster compactifies.

For a quantitative comparison of the density profiles measured in the simulations to our theoretical prediction~\eqref{eq:noa:rho}, we normalize concentrations $c$ with respect to the average concentration $c_0$ in an (unconfined) uniform system at density $\rho_0$. In the mean-field theory (which neglects the repulsive interactions) $c_0$ is given by Eq.~\eqref{eq:c_uni}. In the simulations $c_0$ is reduced due to the finite size of the particles. It can be approximately calculated via Eq.~\eqref{eq:c} by assuming $\rho(\x-\x')\approx \rho_0 \Theta(|\x-\x'|-\sig_\text{eff})$ (with $\Theta(x)$ the Heaviside step-function) in the uniform system. In this way, we obtain
\begin{equation}
 c_0=\frac{\gam \lam}{2D_\text{c}}\rho_0 e^{-\sig_\text{eff}/\lam}
\end{equation}
as corrected normalization constant in the simulations.

\subsubsection{Role of the interface}

\begin{figure}[t]
  \centering
  \includegraphics{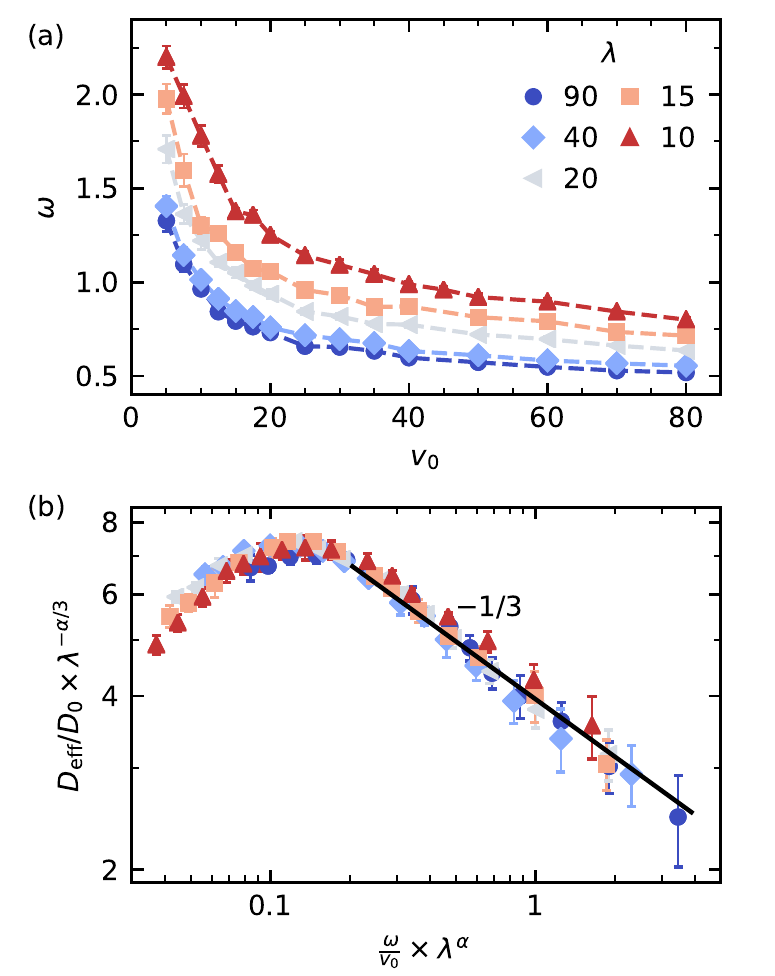}
  \caption{Interfacial width and effective diffusion. (a)~Interfacial width $\omega$ between cluster and gas versus propulsion speed $v_0$ for different decay lengths $\lam$. Error bars show the standard errors. (b)~Fitted effective diffusion coefficients $D_\text{eff}$ for different $\lam$ [same colors as in (a)] as double-logarithmic plot. The data collapses onto a single curve when rescaled by $\lam$ with exponent $\al\simeq0.57$. For broad interfaces, the effective diffusion coefficient becomes independent of $\lam$ and decreases as $D_\text{eff}\sim (\omega/v_0)^{-1/3}$ (solid line). Error bars show standard deviations of fits.}
  \label{fig:circular3}
\end{figure}

For $\lam=10$, speed $v_0=20$, and threshold $\bar{c}=1.4c_0$, Fig.~\ref{fig:circular}(a) compares the density profile of circular clusters obtained in the numerical simulations with the theoretical prediction. We observe that the theoretical density profile decays very sharply at the interface between the regions of velocity zero and $v_0$ (the decay length $\xi_0$ is small), which is in contrast to the broad interface the simulated system exhibits. In Fig.~\ref{fig:circular3}(a) we plot the interfacial width $\omega$ against the propulsion speed $v_0$ in the gas for different decay lengths of the chemical $\lambda$. We observe a notable increase of $\omega$ for smaller $v_0$, which we attribute to enhanced fluctuations. Additionally, the interface becomes broader with decreasing decay length, as already observed in Ref.~\citenum{bauerle18}. This can be explained by fluctuations: For smaller $\lambda$, a particle receives chemical signals from fewer particles and so the total concentration it measures fluctuates stronger. Because of this, a small density fluctuation can trigger a particle at the cluster surface to switch its motility from passive to active (or reverse), which leads to a rougher and thus broader interface. Due to the excluded volume of the particles, additional fluctuations arise in the simulations.

\begin{figure*}[t]
  \centering
  \includegraphics[width=.95\linewidth]{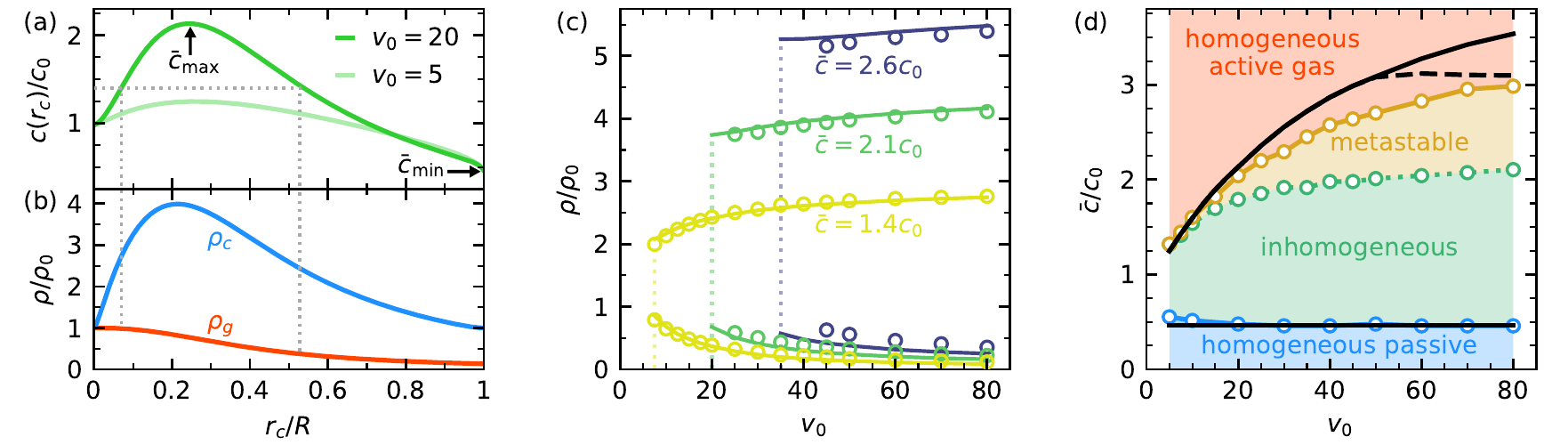}
  \caption{Cluster stability under confinement (at $\lam=10$). (a)~Concentration at cluster boundary $c(\rc)$ [solution of Eq.~\eqref{eq:cBar_rad}] and (b)~coexisting densities as function of cluster size $\rc$ for $v_0=20$. For a given threshold $\bar{c}>c_0$ (dotted line $\bar{c}=1.4c_0$) there are two solutions for $\rc$ with different coexisting densities $\rho_{c,g}$ (in the simulations the solution with larger $\rc$ is selected). The maximum $\bar{c}_\text{max}$ sets the maximal threshold at which passive clusters can exist. For smaller speeds (shown is $v_0=5$) $\bar{c}_\text{max}$ decreases. The value of $c(\rc=R)$ sets the minimum threshold $\bar{c}_\text{min}$ below which the system is entirely passive. (c)~Cluster and gas densities for different $\bar{c}$ as a function $v_0$ from simulations (symbols) and theory (lines). In the simulations, there exists a minimal speed $v_\text{min}$ below which no stable clusters exist. The theory predictions for $v_\text{min}$ [determined by $\bar{c}_\text{max}(v_0)=\bar{c}$] are shown as dotted lines. For $\bar{c}=2.1,2.6c_0$ the system was initialized in a crystalline configuration. (d)~Phase diagram $v_0$--$\bar{c}$ from simulations. Blue: Passive homogeneous system (defined by $\rho_c < 1.02\rho_0$). Green: Passive clusters surrounded by an active gas form from a random starting configuration. Orange: Metastable clusters, which only form when particles are initialized as a crystal. Red: No clusters form, the system is a homogeneous active gas. Upper black line: Maximal threshold $\bar{c}_\text{max}(v_0)$ from theory. Above the dashed line, the theory predicts clusters denser than dense packing. Lower black line: Minimal threshold $\bar{c}_\text{min}$ from Eq.~\eqref{eq:cMin}.}
  \label{fig:circular2}
\end{figure*}

Second, the theory notably overestimates the cluster density and underestimates the gas density. This suggests that in the theory the particle current from the cluster into the gas is significantly smaller than in the simulations. In the mean-field picture, a particle leaves the cluster (it turn from passive to active) when it moves from $r<\rc$ to $r>\rc$, which can only happen  via translational diffusion of the particle itself (the current out of the cluster is $-D_0 \partial_r \rho$). But in the simulations the concentration a particle senses fluctuates (even if the particle would be held fixed) due to the motion of the surrounding particles. Consequently, particles at the cluster boundary constantly change between sensing super- and sub-threshold concentrations (and thus switch between passive and active). When -- due to a concentration fluctuation -- a particle at the surface turns active and its orientation points towards the gas, it will leave the cluster (if it moves far enough away from the surface before it turns passive again due to another fluctuation). Due to this additional fluctuation-induced particle current, the disagreement between theory and simulations is not surprising.

The dominant effect of this process is to induce an additional diffusive current $\vec j_\text{fluc}=-D_\text{fluc}\nabla\rho$ through the interface. Since the density gradient is small away from the interface, we replace $D_0$ by the increased effective diffusion coefficient $D_\text{eff}=D_0+D_\text{fluc}$. Interestingly, we find that it is possible to choose a single $D_\text{eff}$ such that the coexisting densities $\rho_c$ and $\rho_g$ as well as the cluster size $\rc$ show excellent agreement between theory and simulations, see Fig.~\ref{fig:circular}(a). At these parameters, the best agreement (minimum of the sum of squared errors for $\rho_c$ and $\rho_g$) is found for $D_\text{eff}=10.6D_0$, which shows the relevance of the fluctuation-induced current. Moreover, for fixed $v_0$ and $\lam$ this agreement is independent of the threshold value $\bar c$ (in the regime of stable cluster formation), cf. Fig.~\ref{fig:circular}(b,c). We therefore treat $D_\text{eff}(v_0,\lam)$ as an effective parameter depending on decay length $\lam$ and propulsion speed $v_0$, by which we simultaneously fit the theory curves for $\rho_c(\bar c)$ and $\rho_g(\bar c)$ to the simulation data.

The quantity $\omega/v_0$ can be regarded as a measure for the time a particle (which is escaping from the cluster at propulsion speed $v_0$) needs to travel through the interface of width $\omega$ to enter the active gas. When a particle has to spend a longer time in the interfacial region in order to escape, the probability that it turns passive again by a concentration fluctuation (and thus reenters the cluster before reaching the gas) increases. This leads to a diminishing number of particles that successfully leave the cluster and thus to decreasing $D_\text{eff}$. Combining the fitted values $D_\text{eff}(v_0,\lam)$ with the measured interfacial widths $\omega(v_0,\lam)$, we plot $D_\text{eff}$ versus $\omega/v_0$ in Fig.~\ref{fig:circular3}(b). Interestingly, we observe a collapse of the data for
\begin{equation}
  \label{eq:Deff:scal}
  \frac{D_\text{eff}}{D_0} = \lam^{\al/3}f\left(\lam^\al\frac{\omega}{v_0}\right)
\end{equation}
with exponent $\al\simeq0.57$ and scaling function $f(x)$. This function is non-monotonous and decays as $f(x)\sim x^{-1/3}$ for large $x$. In this regime the effective diffusion coefficient becomes independent of the interaction range $\lam$ and decreases, indicating that fluctuations are less important. In the opposite limit of small $x$ corresponding to intrinsically sharp interfaces [with $v_0$ large, cf. Fig.~\ref{fig:circular3}(a)] the fluctuations captured by the effective diffusion coefficient also become less important.

\subsubsection{Phase Diagrams}

Employing the obtained $D_\text{eff}$ values, we now construct the theoretical $v_0$--$\rho$ and $v_0$--$\bar{c}$ phase diagrams for circular clusters, which we quantitatively compare to the simulation results.

We first note that the theory curve $c(\rc)$ [Eq.~\eqref{eq:cBar_rad}] for the concentration at the cluster boundary exhibits a maximum $\bar{c}_\text{max}$ and thus, in principle, allows for two solutions $(r_c,\rho_c,\rho_g)$ for given $\bar{c}$, $v_0$ and $\lam$, cf. Fig.~\ref{fig:circular2}(a,b). However, the solution with the smaller $\rc$ is unstable: Increasing $\rc$, the rim concentration becomes larger than the threshold and thus particles at the rim turn passive and the cluster grows. At the larger solution the reverse holds, increasing the cluster size the rim concentration is below the threshold and rim particle turn active, thus stabilizing the cluster. Consequently, in the simulations always the solution with larger cluster size $\rc$ is observed. For threshold values higher than $\bar{c}_\text{max}$ no solution exists, which implies that the system is an active gas. This upper limit diminishes with decreasing speed $v_0$, see Fig.~\ref{fig:circular2}(a). Hence (as for the planar interface in Sec.~\ref{sec:planar}), for given $\bar c$ a phase-separated state exists only above a minimal speed $v_\text{min}$ [determined by $\bar{c}_\text{max}(v_0)=\bar{c}$]. This is confirmed plotting the coexisting densities as a function of propulsion speed in Fig.~\ref{fig:circular2}(c), which again shows very good agreement with the theoretical predictions.

In Fig.~\ref{fig:circular2}(d), we finally plot the phase diagram for circular clusters [cf. Fig.~\ref{fig:planar}(b)]. In the inhomogeneous regime for small thresholds $\bar c$, in the simulations we observe an instantaneous formation of a dense cluster starting from a disordered initial configuration [green region in Fig.~\ref{fig:circular2}(d), Supplemental Video 1]. Increasing $\bar{c}$, we observe a transition to a nucleation-like regime, where clusters only form after a waiting time when a sufficiently strong density fluctuation has occurred [orange region in Fig.~\ref{fig:circular2}(d), Supplemental Video 2]. In order to access the steady state within reasonable time, in this (metastable) regime we initialize the system in a crystalline configuration. The upper threshold up to which we observe a stable phase-separated state in the simulations is very close to the theoretical prediction $\bar{c}_\text{max}(v_0)$, cf. Fig.~\ref{fig:circular2}(d). Figure~\ref{fig:circular2}(c) shows that also at thresholds in the metastable regime the coexisting densities still agree well with the theory. The remaining discrepancies probably arise due to fluctuations and the influence of the excluded volume, which are both not included in the theory. As there is no upper limit for the cluster density in the theory, at very high speeds it can rise above that of dense packing (corresponding to $\rho_c\approx6.66\rho_0$). At the highest speeds and thresholds we have probed in the simulations, clusters reach densities of $\sim 90\%$ of dense packing. Furthermore, also the prediction for the lower limit $\bar{c}_\text{min}$ below which the system is homogeneous and passive [Eq.~\eqref{eq:cMin}] is in good agreement with the simulations.


\begin{figure*}[t]
  \centering
  \includegraphics[width=.95\linewidth]{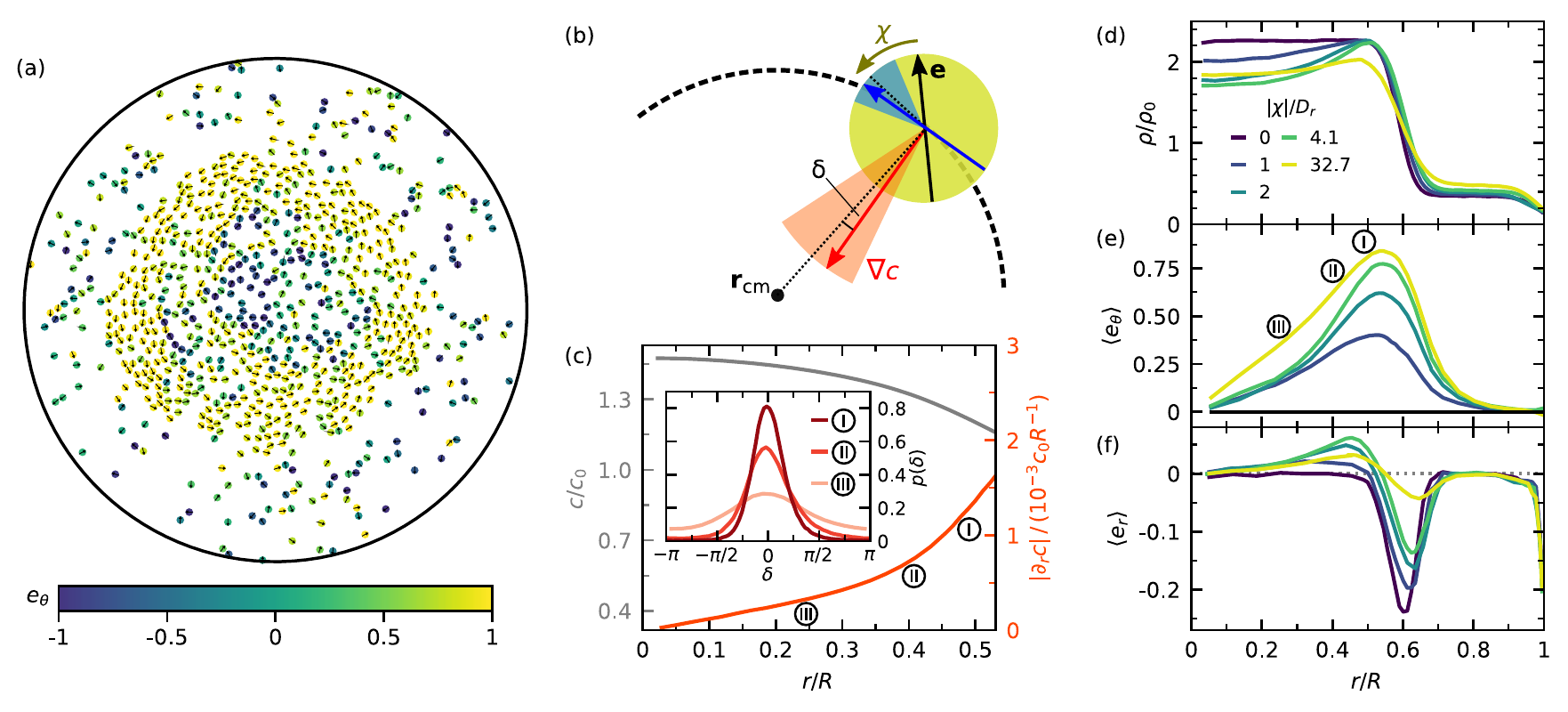}
  \caption{Vortex clusters [$\lam=20$, $\bar c=c_0$, $v^>=30$,  $v^<=\frac{1}{8}v^>$]. (a)~Snapshot of the system with coupling strength $\chi=-32.7$. Particles are colored by their angular alignment, $e_\theta=\vec{e} \cdot \vec{e}_\theta$. (b)~Schematic illustration of the alignment mechanism. On average the concentration gradient $\nabla c$ (red) points towards the particles' center-of-mass $\vec{r}_\text{cm}$. The torque of strength $\chi$ (brown) aligns the particle orientation $\vec{e}$ perpendicular to $\nabla c$ (blue), inducing circular particle motion around $\vec{r}_\text{cm}$ (dashed line). The direction of $\nabla c$ fluctuates around $-\vec{e}_r$ (red cone, angle $\delta$), causing the preferred orientation to fluctuate as well (blue cone). (c)~Radial profile of concentration $c$ and magnitude of radial component of its gradient $|\partial_r c|$ inside the cluster for $\chi=-32.7$. Inset: Distribution of angle $\delta$ for different distances $r$ from center of mass (indicated by Roman numerals). (d)~Radial profiles of density $\rho$, (e)~average angular orientation $\mean{e_\theta}$, and (f)~radial orientation $\mean{e_r}$ for different $\chi$ [labels in (e) refer to (c)].}
  \label{fig:vortices1}
\end{figure*}

\section{Vortex clusters}
\label{sec:vortex}

We now include the coupling of the particle orientations to the instantaneous direction of the concentration gradient, $\vec s_i=\nabla c/|\nabla c|$ evaluated at the particle position $\x_i$ [cf. Fig.~\ref{fig:sketch}(b)]. In particular, we set $\chi^<=\chi\neq 0$, $\chi^>=0$. Thus, according to Eq.~\eqref{eq:eom}, particles with $c_i>\bar{c}$ experience a torque aligning them perpendicular to $\vec s_i$. Moreover, we need to consider finite speeds $v^<>0$ inside the cluster to induce an angular particle current.

\subsection{Simulations}

We perform simulations as described in Sec.~\ref{seq:noa:simulations} with additional orientational coupling. Throughout this section, we set $\lam=20$, $\bar c=c_0$, and $v^>=30$. We employ a slightly smaller confinement radius $R=36.7$, implying a packing fraction $\phi=0.182$ ($\rho_0=0.189$). Again, we observe the formation of a single cluster with reduced speed $v^<$ surrounded by a gas with higher speed $v^>$. Additionally, the orientational coupling induces a angular particle current inside the cluster [see Fig.~\ref{fig:vortices1}(a) for $\chi=-32.7\Dr$ and $v^<=\frac{1}{8}v^>$].

The emergence of this vortex structure can be understood as follows: On average $\nabla c$ points towards the center of the cluster, cf. Fig.~\ref{fig:vortices1}(b,c). Due to the aligning torque particles orient perpendicular to the gradient and thus swim along regions of constant $c$, which correspond to circles around the cluster's center, see Fig.~\ref{fig:vortices1}(b). Depending on the sign of $\chi$, particles swim clockwise ($\chi>0$) or counter-clockwise ($\chi<0$).

In order to quantify the orientational order in the system, we change to polar coordinates with the particles' center of mass at the origin and unit vectors $\vec{e}_r$ and $\vec{e}_\theta$. We calculate the average radial and angular component of the particle orientations
\begin{equation}
  \mean{e_r}=\mean{\vec{e}\cdot \vec{e}_r}, \quad \mean{e_\theta}=\mean{\vec{e}\cdot \vec{e}_\theta}
\end{equation}
as a function of the distance $r$ to the center of mass. Since rotational diffusion is counteracting the aligning torque, $\mean{e_\theta}$ increases with rising torque strength $\chi$ [see Fig.~\ref{fig:vortices1}(e)]. At the inner side of the cluster-gas interface, $\mean{e_\theta}$ has a maximum and decays both into the gas (where $\chi=0$) and towards the cluster's center. The latter arises from fluctuations of the concentration field: as the magnitude of $\mean{\partial_r c}$ diminishes towards the center, the variance of fluctuations of $\vec s_i$ around its mean $\mean{\vec s_i}=-\vec{e}_r$ increases, cf. Fig.~\ref{fig:vortices1}(c). In turn, the direction along which particles are aligned by the torque is fluctuating stronger, leading to the drop of $\mean{e_\theta}$. Regions of increased angular alignment have a lower outgoing particle current in radial direction. Therefore, the cluster density is not homogeneous anymore but exhibits a peak close to the position of maximal $\mean{e_\theta}$, cf. Fig.~\ref{fig:vortices1}(d).

The $\mean{e_r}$-profile exhibits two peaks: a strong negative peak at the interface between the regions of small and high $v$ (particle orientations on average point into the cluster) and a weaker positive peak on the inner side of the density maximum [see Fig.~\ref{fig:vortices1}(f)]. The negative peak diminishes with rising torque strength, which leads to a stronger particle current out of the cluster and thus to an increased gas density, cf. Fig.~\ref{fig:vortices1}(d).

\subsection{Theory}

For completeness, we generalize the analysis of circular clusters to orientational coupling and to provide the explicit expressions. We set $\chi^<=\chi$ and $\chi^>=0$ and use the mean direction $\vec s=-\vec{e}_r$ pointing towards the origin. The polarization $\vec p(r)=p_r \vec e_r+p_\theta(r)\vec{e}_\theta$ now has a radial and angular component. In the steady state, the radial component $j_r$ of the current still vanishes, $j_r=0$, implying $\partial_r\rho=(v/D_0)p_r$. In contrast, the angular current is now non-zero and given by $j_\theta=v p_\theta$.

For the polarization equation~\eqref{eq:p}, in the inner region we need the matrix $\vec M^<=\vec M_0+\vec M_\chi$ with $\vec M_0$ from Eq.~\eqref{eq:noa:M} plus the matrix
\begin{multline}
  \vec M_\chi = \frac{\chi v}{16\Dr}\left(-3p_\theta\vec e_r\vec e_r + p_r\vec e_r\vec e_\theta + 3p_r\vec e_\theta\vec e_r + p_\theta\vec e_\theta\vec e_\theta\right)
\end{multline}
due to the coupling of the orientation. We then arrive at the two coupled inhomogeneous modified Bessel equations
\begin{align}
  \label{eq:deq_pR_circular}
  0 ={}& p_r'' + \frac{p_r'}{r} - \frac{p_r}{r^2} - \frac{p_r}{\xi_\chi^2} - \frac{3\chi v}{v_\ast^2+v^2}\left(p_\theta'+\frac{p_\theta}{r}\right), \\
 \begin{split} \label{eq:deq_pTheta_circular}
 0 ={}& p_\theta''+ \frac{p_\theta'}{r} -\frac{p_\theta}{r^2} -\frac{p_\theta}{\tilde \xi_\chi^2} + \frac{3\chi v}{v_\ast^2+v^2} \left( p_r'+\frac{p_r}{r}\right) \\
 &\quad \quad \quad \quad \quad \quad \quad \quad \,\, -\frac{\chi}{2D_0[1+(v/v_\ast)^2]}\rho,
 \end{split}
\end{align}
with radial
\begin{equation}
  \xi_\chi \equiv \ell\left[\frac{1+(v/v_\ast)^2}{1+8(v/v_\ast)^2+\tfrac{1}{8}(\chi/\Dr)^2}\right]^{1/2}
\end{equation}
and angular decay lengths
\begin{equation}
  \tilde \xi_\chi \equiv \ell\left[\frac{1+(v/v_\ast)^2}{1+\tfrac{1}{8}(\chi/\Dr)^2}\right]^{1/2},
\end{equation}
which generalize the decay length $\xi_0$ [Eq.~\eqref{eq:xi}] obtained in the absence of orientational coupling.

\begin{figure}[b!]
  \centering
  \includegraphics{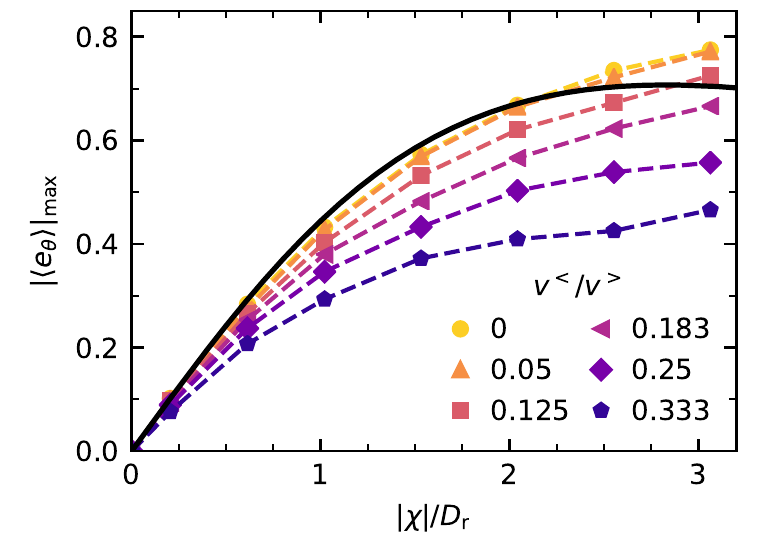}
  \caption{Maximum $|\mean{e_\theta}|_\text{max}$ of the average angular particle orientation $|\mean{e_\theta}|(r)$ versus torque strength $|\chi|$ for different velocities $v^<$ in the cluster. Parameters are $\lam=20$, $v^>=30$, and $\bar c=c_0$ (standard errors are smaller than symbol size). The $v^<$-independent theory prediction based on Eq.~\eqref{eq:p_theta_noD0} is shown in black. For higher $|\chi|$, $|\mean{e_\theta}|_\text{max}$ saturates and the theory line decays to zero (data not shown).}
  \label{fig:vortices2}
\end{figure}

Obtaining a closed analytical solution of Eqs.~\eqref{eq:deq_pR_circular} and \eqref{eq:deq_pTheta_circular} appears to be a formidable task. Nevertheless, one can draw some qualitative insights into the behavior of the system through further simplifications. Here we just give one example for the maximum $\mean{e_\theta}_\text{max}$ of the angular orientation profiles. As simplification, we consider the limit of vanishing translational diffusion. In that case, the density jumps between the constant values $\rho_c$ and $\rho_g$ at $r_c$ and $p_r=0$ everywhere. Then, Eq.~\eqref{eq:deq_pTheta_circular} implies $p_\theta=0$ for $r>r_c$ and 
\begin{equation}
  p_\theta=-\frac{1}{2}\frac{\chi/\Dr}{1+\chi^2/(8\Dr^2)}\rho_c
  \label{eq:p_theta_noD0}
\end{equation}
in the cluster, resulting from the competition of alignment and rotational diffusion. Equation~\eqref{eq:p_theta_noD0} implies an average angular orientation $\mean{e_\theta}=p_\theta/\rho_c$. Comparing this prediction to the maximum $\mean{e_\theta}_\text{max}$ of the angular orientation profiles from the simulations, we find good agreement for small $v^<$ and $\chi$, cf. Fig.~\ref{fig:vortices2}. With increasing $v^<$, (density and concentration) fluctuations become more important, leading to stronger deviations from the mean-field prediction. For $\chi\gtrsim 3 \Dr$, $\mean{e_\theta}_\text{max}$ saturates in the simulations, but the theory prediction decays to zero for $\chi\to \infty$ (data not shown). In order to capture the high $\chi$-behavior, we probably would need to take into account higher moments of the probability density.


\section{Conclusions}

Taking into account a discontinuous motility leads to a novel paradigm for aggregation in scalar active matter different from chemotaxis and the motility-induced phase separation of active Brownian particles~\cite{cate15}. Particles aggregate into dense domains through a reduction of their motility in response to an event, here passing a predefined threshold of a self-generated field. This constitutes a simple communication mechanism at the microscale through which collective behavior can be initiated. It is exploited in nature through quorum-sensing of signaling molecules exuded by all members of a population and has recently been demonstrated in synthetic active matter~\cite{bauerle18}. Particles switch their motility depending on external cues, which sets them apart from mixtures of particles with different but fixed motility~\cite{ni14,kummel15,sten15,sho15,wysocki2016} (or diffusivity~\cite{weber16}). Implementing such communication beyond direct physical (steric, hydrodynamic, etc.) forces allows to ``program'' responses and thus to tailor collective behavior. In particular, the aggregation of building blocks is a crucial step in the autonomous assembly of ordered, hierarchical structures. Controlling the self-assembly of active matter is an important step towards microscopic ``machines of machines''~\cite{needleman17} and will become important for the emerging field of microrobots performing tasks collectively.

We have derived and investigated the coupled linear dynamic equations for density and polarization, whereby the propulsion speed $v(c)$ is now a function of the concentration $c$ and the concentration is a linear functional of the density [Eq.~\eqref{eq:c}]. Specifically, we have studied a piece-wise constant function $v(c)$ that drops to zero above a threshold $\bar c$. The resulting spatially inhomogeneous steady state is reminiscent of the phase coexistence of an active gas with a dense liquid separated by an interfacial region. As for active Brownian particles, coexistence is determined by the balance of diffusive and active particle current. The later is due to a non-zero polarization induced by the interface, which decays into the bulk on a length scale $\xi_0$ [Eq.~\eqref{eq:xi}]. Most notable, however, is the absence of a critical point for discontinuous motilities. Introducing as single fit parameter an effective diffusion coefficient accounting for the fluctuations of the concentration field within the interface, excellent agreement with particle-based numerical simulations has been demonstrated. The effective diffusion coefficient is independent of the threshold and obeys a scaling form [cf. Eq.~\eqref{eq:Deff:scal} and Fig.~\ref{fig:circular3}]. This success of a mean-field theory might indeed be attributed to the absence of a critical point, which is characterized by diverging fluctuations and, consequently, the break-down of mean-field approximations in its vicinity.

The theoretical framework presented here is not restricted to the exponential kernel in Eq.~\eqref{eq:c} for the chemical concentration field and other forms, \emph{e.g.} generated through light, might be considered. Moreover, while for clarity we have focused on a single circular cluster in confinement, the aggregation is not an artifact of the confinement but can also be observed in simulations employing periodic boundary conditions (Supplemental Video 3). We have also discussed a novel route to the formation of vortex clusters, in which particles respond to the local field gradient through a torque. Such torques can be implemented experimentally for colloidal Janus particles~\cite{lozano16}. Again, vortex formation is not linked to confinement but occurs in extended systems using periodic boundary conditions (Supplemental Video 4), which has been observed experimentally for magnetic rollers~\cite{kokot18}.

In this work we have treated only steady state profiles. It would be interesting to investigate in more detail the nature of density fluctuations and the temporal evolution of interfaces, which connects active matter to the classical Stefan problem~\cite{gupta}.


\begin{acknowledgments}
  We gratefully acknowledge financial support by the Deutsche Forschungsgemeinschaft (DFG) through the Graduate School of Excellence ``Materials Science in Mainz'' (GSC 266) and the priority program SPP 1726 (grant no. 254473714). Numerical computations were carried out on the MOGON2 Cluster at ZDV Mainz.
\end{acknowledgments}

\appendix

\section{Derivation of effective hydrodynamic equation}
\label{sec:hydro}

In order to derive the dynamic equations \eqref{eq:rho} and \eqref{eq:p} for the density and polarization, respectively, we write Eq.~\eqref{eq:psi} in cartesian coordinates $(x,y)$ and measure the particle orientation $\vhi$ with respect to the $x$-axis. We follow Ref.~\citenum{bertin06} and calculate the evolution equations for the moments
\begin{gather}
  \psi_n(x,y)=\IInt{\vhi}{0}{2\pi} \cos(n\vhi)\psi(x,y,\vhi), \\
  \phi_n(x,y)=\IInt{\vhi}{0}{2\pi} \sin(n\vhi)\psi(x,y,\vhi)
\end{gather}
yielding a hierarchy of equations. For $n=0$ we obtain
\begin{equation}
 \label{eq:psi0}
 \pd{\psi_0}{t}=-\pd{}{x}(v\psi_1)-\pd{}{y}(v\phi_1)+D_0\nabla^2\psi_0.
\end{equation}
For $n=1$ we have
\begin{multline}
 \label{eq:psi1}
 \pd{\psi_1}{t}=-\frac{1}{2}\left[\pd{}{x}(v\psi_0)+\pd{}{x}(v\psi_2)+\pd{}{y}(v\phi_2)\right] \\ + D_0\nabla^2\psi_1 
  + \frac{1}{2}\chi \left[s_y(\psi_2-\psi_0)-s_x\psi_2\right]-\Dr\psi_1
\end{multline}
and
\begin{multline}
 \label{eq:phi1}
 \pd{\phi_1}{t}=-\frac{1}{2}\left[\pd{}{y}(v\psi_0)+\pd{}{x}(v\phi_2)-\pd{}{y}(v\psi_2)\right] \\ + D_0\nabla^2\phi_1 
  + \frac{1}{2}\chi \left[s_y\phi_2+s_x(\psi_2+\psi_0)\right]-\Dr\phi_1.
\end{multline}
By setting $\psi_3=\phi_3=0$ and neglecting time and spatial derivatives of $\psi_2$ and $\phi_2$, at the next order $n=2$ we obtain
\begin{equation}
 \label{eq:psi2}
 \psi_2=\frac{1}{4\Dr}\left[-\frac{1}{2}\pd{}{x}(v\psi_1)+\frac{1}{2}\pd{}{y}(v\phi_1)-\chi(s_x\phi_1+s_y\psi_1)\right]
\end{equation}
and
\begin{equation}
 \label{eq:phi2}
 \phi_2=\frac{1}{4\Dr}\left[-\frac{1}{2}\pd{}{x}(v\phi_1)-\frac{1}{2}\pd{}{y}(v\psi_1)-\chi(s_y\phi_1-s_x\psi_1)\right],
\end{equation}
which now only couple to the $n=1$ modes. Finally, plugging Eqs.~(\ref{eq:psi2},\ref{eq:phi2}) into Eqs.~(\ref{eq:psi1},\ref{eq:phi1}) and identifying density $\rho\equiv\psi_0$ and polarization $\vec p\equiv(\psi_1,\phi_1)$ we arrive at Eqs.~\eqref{eq:rho} and \eqref{eq:p} from the main text. Note that for the divergence of a tensor, we use the convention $(\nabla\cdot \vec M)_j=\partial_i M_{ij}$. 


%

\end{document}